\documentclass[
    ,final            
  ]
  {aipproc}

\layoutstyle{6x9}

\newcommand{\Ep}{\ensuremath{E_{peak}}}

\begin{document}

\title{RHESSI Spectral Fits of \textit{Swift} GRBs}



\classification{98.70.Rz; 95.85.Pw}
\keywords      {gamma-rays: bursts}

\author{Eric C. Bellm}{
  address={UC Berkeley Space Sciences Laboratory, 7 Gauss Way,
  Berkeley, CA 94720-7450, USA}
  ,altaddress={Department of Physics, UC Berkeley} 
}

\author{Mark E. Bandstra}{
  address={UC Berkeley Space Sciences Laboratory, 7 Gauss Way,
  Berkeley, CA 94720-7450, USA}
  ,altaddress={Department of Physics, UC Berkeley} 
}

\author{Steven E. Boggs}{
  address={UC Berkeley Space Sciences Laboratory, 7 Gauss Way,
  Berkeley, CA 94720-7450, USA}
  ,altaddress={Department of Physics, UC Berkeley} 
}

\author{Wojtek Hajdas}{
  address={Paul Scherrer Institute, Villigen PSI, Switzerland}
}

\author{Kevin Hurley}{
  address={UC Berkeley Space Sciences Laboratory, 7 Gauss Way,
  Berkeley, CA 94720-7450, USA}
}

\author{David M. Smith}{
  address={Santa Cruz Institute for Particle Physics, Santa Cruz, CA, USA}
}

\author{Claudia Wigger}{
  address={Paul Scherrer Institute, Villigen PSI, Switzerland}
}

\begin{abstract}
One of the challenges of the \textit{Swift} era has been accurately determining
\Ep\ for the prompt GRB emission.  RHESSI, which is sensitive from
30 keV to 17 MeV, can extend spectral coverage above the \textit{Swift}-BAT bandpass.
Using the public \textit{Swift} data, we present results of joint spectral fits for 26
bursts co-observed by RHESSI and \textit{Swift}-BAT through May 2007.  
We compare these fits to estimates
of \Ep\ which rely on BAT data alone.  A Bayesian \Ep\
estimator gives better correspondence with our measured results than an
estimator relying on correlations with the \textit{Swift} power law indices.
\end{abstract}

\maketitle


\section{GRB Prompt Emission Spectroscopy with RHESSI}

The Reuven Ramaty High-Energy Solar Spectroscopic Imager (RHESSI)
\citep{lin02} is a dedicated solar observatory.  
RHESSI's nine germanium detectors are sensitive
from 30 keV to 17 MeV, with excellent resolution in energy (1-5 keV) and
time (1 binary $\mu$s) \citep{smit02}.  Each of the nine coaxial
detectors is electronically segmented into front and rear segments.
Because the detectors are unshielded, RHESSI observes emission from
astrophysical sources like GRBs with a $\sim$2$\pi$ field of view.

We perform Monte Carlo simulations using the MGEANT package \citep{stur00} 
to determine RHESSI's response 
to off-axis sources like GRBs.  RHESSI's response varies with off axis angle, 
so we create responses every 15 degrees.  For each response, we simulate 
monoenergetic photons in 192 logarithmic energy bins ranging from
30 keV -- 30 MeV.  Since RHESSI's per-detector response also varies during
the spacecraft's four second spin period, 
we bin the annular response in six azimuthal bins and weight these bins by the 
total burst lightcurve to create the final response.

The RHESSI data are extracted in SSW-IDL.  We fit and subtract a 
time-varying background, allowing for possible periodic modulation with 
the spin period.  We perform spectral fitting in ISIS \citep{citeisis}, 
a forward-fitting package analogous to 
XSPEC\footnote{http://heasarc.gsfc.nasa.gov/docs/xanadu/xspec/},
which is extensively programmable and allows 
computation of rigorous fluence error estimates via exploration of 
the parameter space.

Energetic charged particles over time have caused radiation damage to 
RHESSI's germanium detectors.  This radiation damage causes broadened 
spectral lines due to hole trapping and a general loss of active volume.  
In this work, we restrict our 
analysis to those detectors which do not exhibit signs of radiation damage.
At the time of \textit{Swift}'s launch, six RHESSI segments were usable for
spectroscopy; by May 2007 only two segments remained undamaged.  
In November 2007, 
RHESSI underwent an annealing procedure to reverse the effects of the
radiation damage.  The anneal restored some lost sensitivity, but 
analysis of future bursts will require more sophisticated modeling of
the remaining effects of radation damage on RHESSI's spectral response.

\section{RHESSI-BAT Joint Fits}

We attempted simultaneous spectral fitting for all RHESSI-observed GRBs
appearing in the first BAT Catalog \citep{saka07}.
Of 46 candidate bursts, 26 had sufficient RHESSI counts for spectral
analysis and produced acceptable joint fits.  We selected analysis time 
intervals manually from the RHESSI lightcurve using a S/N criterion. 
The resulting intervals were usually
shorter than those used in the BAT Catalog.  We generated BAT spectra and
responses for our intervals with the standard analysis 
procedures.\footnote{http://swift.gsfc.nasa.gov/docs/swift/analysis/threads/bat\_threads.html}

Typically, the joint fits did not require a normalization 
offset between the RHESSI and BAT data.  (Of the four bursts that 
needed an offset for acceptable fitting, one was in a period of highly modulated
RHESSI background, and three were after December 2006 when radiation 
damage was becoming severe.) 
For the RHESSI data, we generally fit over the full 30
keV-17 MeV energy band.  For bursts coming from the rear of RHESSI, we
raised the lower energy bound to $\sim$60 keV, as the additional passive
material of the RHESSI cryostat can influence the low energy data.  For
GRB 061007, we omitted the RHESSI data above 3 MeV; there were no
significant counts above that level, but the residuals showed systematic
deviation which biased the fit.

Our results show good correspondence with comparable fits reported by
Konus-Wind and Suzaku-WAM, which also are sensitive in the MeV range
\citep[e.g.,][]{krim06, kgcn061007, sgcn061007, kgcn061121, page07}.
For 16 of the 26
bursts, the joint RHESSI-BAT best fit found additional model parameters
(\Ep\ and/or $\beta$) compared to the BAT-only fit.

We report the results of the joint fits in Table \ref{tab-fitpars}.

\section{Testing \Ep\ Estimators}

\begin{figure}
  \resizebox{\textwidth}{!}{\includegraphics{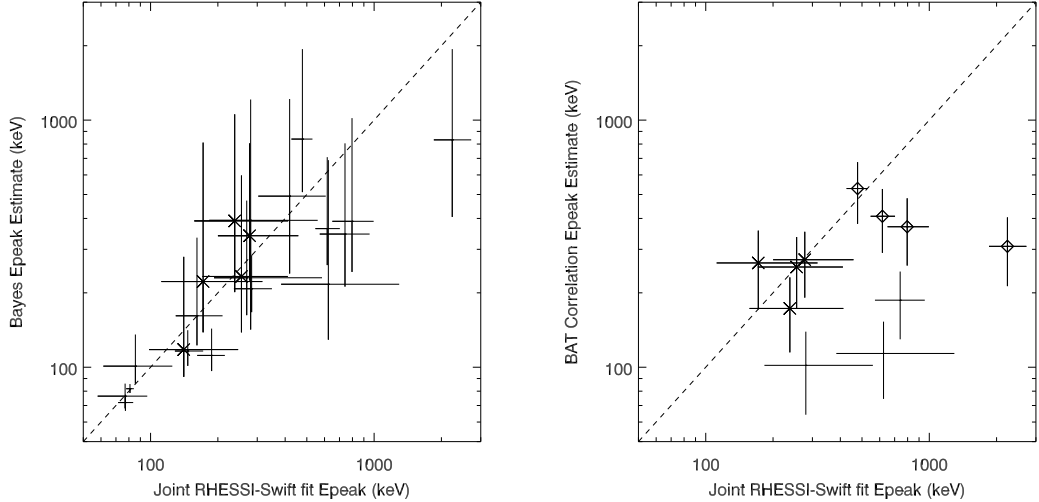}}
  \caption{Comparison of the measured RHESSI+BAT \Ep\ values with those
  predicted by \citep{butl07} (left) and those found with the 
  correlation of \citep{zhan07} (right).
  Points marked with a cross have fit $\beta > -2$, and hence \Ep\ is only a
  formal value.  Points in the right plot marked with a diamond have
  BAT-only power
  law indices outside the range ($-2.3 < \alpha < -1.2$) likely to yield 
  accurate predictions of \Ep\ \citep{zhan07b}.  
  The overplot lines represent equality of the measured and predicted values.}
  \label{fig-cfepeak}
\end{figure}

The peak energy \Ep\ of the prompt GRB spectrum is crucial to determining 
overall burst energetics, and it plays a key role in several proposed 
luminosity indicators.
The narrow passband of \textit{Swift}-BAT prevents determination of \Ep\ for 
many bursts.  Accordingly, a number of attempts have been made to infer
\Ep\ from the BAT data alone.  
\citet{butl07} used a Bayesian fit method with priors determined from the 
BATSE catalog to estimate \Ep.  \citet{zhan07} derived an \Ep\ -- BAT 
power law index correlation using hardness ratios (see also
\citep{zhan07b}).
In Figure \ref{fig-cfepeak}, we compare the predictions of these models 
to our joint fit results.

The Bayes model shows good correspondence with the measured values.  
Above $\sim$600 keV, the predicted values of \Ep\ tend to be low, 
although their error bars reach near the measured values.
This deviation is exaggerated somewhat, as our data
are for shorter, more intense (and typically harder) burst
intervals than used in \citep{butl07}.

Comparison of the Zhang et al. correlation-predicted \Ep\ to our
measured RHESSI-\textit{Swift} values shows that
this correlation appears to systematically underpredict the 
measured value of \Ep, especially at high energy.

\begin{theacknowledgments}
This work was supported by the \textit{Swift} AO-3 GI grant NNX08AE86G.
\end{theacknowledgments}


\bibliographystyle{aipproc}   

\bibliography{grb}

\begin{table}
\begin{tabular}{lccrrrrr}
\hline

\tablehead{1}{c}{c}{GRB} &
\tablehead{1}{c}{c}{Analysis Interval \\ (sec)} &
\tablehead{1}{c}{c}{Best Fit\\Model} &
\tablehead{1}{c}{c}{$\alpha$} &
\tablehead{1}{c}{c}{\Ep \\ (keV)} &
\tablehead{1}{c}{c}{$\beta$} &
\tablehead{1}{c}{c}{Fluence
\\ (10$^{-6}$ ergs/cm$^2$)} &
\tablehead{1}{c}{c}{$\chi^2_\nu$} \\

\hline

041223 & ~~0.0 -- 124.0 & CPL & -0.97$^{+0.05}_{-0.04}$ &  617.$^{ +86.}_{ -70.}$ & $^{}_{}$ &  87.8$^{ +8.1}_{ -7.1}$ &  1.21 \\
041224 & ~-3.7 -- ~36.3 & Band & -0.77$^{+0.60}_{-0.32}$ &   77.$^{ +20.}_{ -19.}$ & -2.10$^{+0.16}_{-0.24}$ &  12.9$^{ +3.8}_{ -3.2}$ &  0.93 \\
050124 & ~-2.2 -- ~~3.8 & Band & -0.65$^{+0.72}_{-0.45}$ &   86.$^{ +40.}_{ -24.}$ & -2.12$^{+0.20}_{-0.50}$ &   3.3$^{ +1.3}_{ -1.2}$ &  0.81 \\
050219B & -29.3 -- ~10.7 & Band & -1.10$^{+0.12}_{-0.10}$ &  188.$^{ +27.}_{ -26.}$ & -2.62$^{+0.29}_{-0.84}$ &  37.1$^{ +5.1}_{ -4.7}$ &  0.99 \\
050326 & ~-1.1 -- ~24.9 & Band & -1.01$^{+0.13}_{-0.11}$ &  277.$^{+181.}_{ -77.}$ & -1.73$^{+0.12}_{-0.19}$ &  61.7$^{+19.1}_{-16.0}$ &  0.72 \\
050525A & ~~0.1 -- ~12.9 & Band & -1.02$^{+0.11}_{-0.10}$ &   81.$^{  +3.}_{  -3.}$ & -3.12$^{+0.25}_{-0.50}$ &  21.3$^{ +1.2}_{ -1.1}$ &  0.90 \\
050713B & ~-1.6 -- ~27.4 & PL & -1.40$^{+0.04}_{-0.04}$ & $^{}_{}$ & $^{}_{}$ &  43.0$^{ +4.8}_{ -4.7}$ &  1.03 \\
050717 & ~-0.2 -- ~35.8 & CPL & -1.13$^{+0.04}_{-0.04}$ & 2237.$^{+483.}_{-386.}$ & $^{}_{}$ &  65.7$^{ +7.2}_{ -6.9}$ &  0.92 \\
050802 & ~-3.3 -- ~22.7 & PL & -1.65$^{+0.06}_{-0.07}$ & $^{}_{}$ & $^{}_{}$ &  15.8$^{ +3.6}_{ -3.2}$ &  1.02 \\
050820B & ~~6.8 -- ~12.8 & CPL & -0.57$^{+0.17}_{-0.16}$ &  147.$^{ +25.}_{ -18.}$ & $^{}_{}$ &   2.6$^{ +0.3}_{ -0.3}$ &  0.96 \\
051111 & -14.5 -- ~11.5 & Band & -1.04$^{+0.19}_{-0.14}$ &  255.$^{+156.}_{ -84.}$ & -1.98$^{+0.17}_{-0.46}$ &  20.0$^{ +4.3}_{ -4.6}$ &  0.89 \\
051221A & ~~0.2 -- ~~2.3 & Band & -1.26$^{+0.17}_{-0.13}$ &  238.$^{+175.}_{ -81.}$ & -1.98$^{+0.11}_{-0.21}$ &   4.8$^{ +0.8}_{ -0.8}$ &  1.35 \\
060110 & ~-0.8 -- ~11.2 & PL & -1.60$^{+0.05}_{-0.05}$ & $^{}_{}$ & $^{}_{}$ &   9.7$^{ +1.8}_{ -1.6}$ &  0.90 \\
060117 & ~-0.6 -- ~17.4 & CPL & -1.58$^{+0.10}_{-0.09}$ &   77.$^{  +6.}_{  -6.}$ & $^{}_{}$ &  26.8$^{ +1.4}_{ -1.2}$ &  1.06 \\
060418 & ~-5.2 -- ~30.8 & CPL & -1.52$^{+0.07}_{-0.07}$ &  624.$^{+671.}_{-240.}$ & $^{}_{}$ &  22.4$^{ +7.7}_{ -4.4}$ &  0.92 \\
060421A & ~-0.3 -- ~~3.7 & Band & -0.88$^{+0.33}_{-0.26}$ &  141.$^{+106.}_{ -42.}$ & -1.87$^{+0.20}_{-0.63}$ &   3.6$^{ +2.1}_{ -1.9}$ &  0.82 \\
060501 & ~-2.9 -- ~10.1 & PL & -1.37$^{+0.05}_{-0.06}$ & $^{}_{}$ & $^{}_{}$ &  21.9$^{ +4.5}_{ -4.2}$ &  1.04 \\
060502A & -15.1 -- ~~9.9 & Band & -0.87$^{+0.35}_{-0.24}$ &  172.$^{+144.}_{ -60.}$ & -1.98$^{+0.26}_{-1.67}$ &  11.8$^{ +7.0}_{ -5.9}$ &  1.04 \\
060614 & ~-1.5 -- ~~3.5 & CPL & -1.56$^{+0.12}_{-0.11}$ &  280.$^{+279.}_{ -97.}$ & $^{}_{}$ &   6.7$^{ +6.7}_{ -1.2}$ &  0.95 \\
060908 & ~-3.3 -- ~~1.7 & CPL & -0.68$^{+0.23}_{-0.20}$ &  161.$^{ +49.}_{ -32.}$ & $^{}_{}$ &   2.0$^{ +0.4}_{ -0.3}$ &  0.87 \\
061007 & ~-0.8 -- ~63.2 & Band & -0.85$^{+0.04}_{-0.04}$ &  478.$^{ +52.}_{ -51.}$ & -2.47$^{+0.24}_{-0.42}$ & 244.6$^{+23.2}_{-22.5}$ &  0.84 \\
061121 & ~56.0 -- ~85.0 & Band & -1.31$^{+0.04}_{-0.04}$ &  741.$^{+214.}_{-169.}$ & -2.37$^{+0.32}_{-1.21}$ &  64.7$^{ +8.1}_{ -8.4}$ &  1.02 \\
061126 & ~-4.4 -- ~22.6 & CPL & -1.06$^{+0.06}_{-0.05}$ &  796.$^{+199.}_{-146.}$ & $^{}_{}$ &  34.0$^{ +5.3}_{ -4.4}$ &  0.92 \\
061222A & ~81.9 -- ~87.9 & CPL & -0.80$^{+0.17}_{-0.17}$ &  269.$^{+316.}_{ -77.}$ & $^{}_{}$ &   7.5$^{ +6.6}_{ -1.8}$ &  0.91 \\
070220 & ~~0.1 -- ~32.1 & CPL & -1.24$^{+0.08}_{-0.07}$ &  419.$^{+189.}_{-116.}$ & $^{}_{}$ &  20.3$^{ +4.9}_{ -3.6}$ &  0.95 \\
070508 & ~~2.2 -- ~18.2 & CPL & -1.07$^{+0.07}_{-0.06}$ &  283.$^{ +66.}_{ -47.}$ & $^{}_{}$ &  42.3$^{ +6.4}_{ -4.9}$ &  0.66 \\

\hline
\caption{
Best joint fit parameters.  Spectral models considered are a Band function
\citep{band93}, cutoff power law (CPL), and simple power law (PL).  
Errors are quoted at the 90\%
confidence level. Fluence is in the 15 keV--10 MeV band, and
analysis times are relative to the BAT trigger time.
}
\label{tab-fitpars}

\end{tabular}
\end{table}

\end{document}